\documentclass{article}
\usepackage[utf8]{inputenc}
\usepackage{authblk}
\usepackage[english]{babel}

\usepackage[letterpaper,top=2cm,bottom=2cm,left=3cm,right=3cm,marginparwidth=1.75cm]{geometry}

\usepackage{amsmath}
\usepackage{graphicx}
\usepackage[colorlinks=true, allcolors=blue]{hyperref}
\def\simpropto{\lower.2ex\hbox{$\; \buildrel \propto \over \sim \;$}}
\def\ltsim{\lower.5ex\hbox{$\; \buildrel < \over \sim \;$}}
\def\gtsim{\lower.5ex\hbox{$\; \buildrel > \over \sim \;$}}

\title{\bf The Future of   Primordial Black Holes:\\
Open Questions and Roadmap}

\author[1,2]{\bf Antonio Riotto}
\author[3,4,5]{\bf Joe Silk}
\affil[1]{D\'epartement de Physique Th\'eorique,
Universit\'e de Gen\`eve, \\24 quai Ansermet, CH-1211 Gen\`eve 4, Switzerland}
\affil[2]{Gravitational Wave Science Center (GWSC), Universit\'e de Gen\`eve,\\ CH-1211 Geneva, Switzerland}
\affil[3]{Institut d’Astrophysique, UMR 7095 CNRS, Sorbonne Universit\'e, \\98bis Bd Arago, 75014 Paris, France}
\affil[4]{Department of Physics and Astronomy, The Johns Hopkins University,\\ Baltimore MD 21218, USA}
\affil[5]{Beecroft Institute of Particle Astrophysics and Cosmology, Department of Physics, University of Oxford, Oxford OX1 3RH, UK}

\begin{document}
\maketitle
\begin{abstract}
\noindent
We discuss  some of the the open questions and the roadmap  in the physics of primordial black holes. Black holes  are the only dark matter candidate that is known to actually exit. Their conjectured primordial role is admittedly based on hypothesis rather than fact, most straightforwardly as a simple extension to the standard models of inflation, or even,  in homage to quantum physics,  more controversially  via a slowing-down of Hawking evaporation. Regardless of one's stance on the theoretical basis for their existence, the possibility of  primordial black holes playing a novel role in dark matter physics and gravitational wave astronomy opens up a rich astrophysical phenomenology that we lay out in this brief overview.
\end{abstract}

\section{Introduction}
Primordial black holes (PBHs) provide  an attractive way to resolve  the dark matter problem. Unlike essentially all other dark matter particle candidates, BHs actually exist, and there are several robust prospects for determining whether there is a primordial component over a broad mass range.  Theoretically PBHs may populate the universe in a humongous range of masses from $10^{-20}\, M_\odot $
to $10^{20}\, M_\odot $. They could be the dark matter for masses around $10^{-12}\, M_\odot $.
Moreover  even if PBHs are dark matter-subdominant, , they could be  responsible for some of the currently  observed mergers of BHs in the solar mass range or   they may play key roles as   rare IMBH (Intermediate Mass BHs in the range $(10^3-10^6) M_\odot$ that may seed both SMBH (Supermassive Black Holes observed in the mass range $(10^6-10^{10})  M_\odot$ and massive galaxies in the early universe.

PBHs  formed very early and hence may be ubiquitous at the highest redshifts accessible by direct imaging, that is $z \ltsim 100,$
when there was little astrophysical competition. This provides us with the possibility of finding truly unique high redshift signatures  of PBHs, for example via gravity wave interferometers such as Einstein Telescope  and Cosmic Explorer that will probe $z\ltsim 20$, for potential sources of EMRI signals.  

Well before these telescopes are in action,  there will be  21cm dark ages telescopes  destined for the lunar far side, such as LUSEE-Night and DSL to probe PBH accretion imprints on diffuse hydrogen absorption against the CMB to $z\ltsim 100$,  both scheduled for deployment in 2026. And on a similar time-scale we anticipate LIGO-class gravity wave telescopes to make significant inroads on the possible existence of  solar mass black holes as well as  black holes populating the "forbidden" region in the pair instability gap, at  masses $\gtsim 50\, M_\odot.$ Such observations would provide further evidence for the existence of PBHs. 

We review some of these arguments  in this chapter, with particular attention to the open questions, challenges and future opportunities.

\section{Some open questions}

How well do we know the PBH abundance? And over what mass range? How well can we ascertain  the PBH clustering properties and therefore the GW signal from binary mergers via LIGO and future terrestrial GW interferometers? And can we use the vicinities of SMBH amd IMBH, most likely rich in dark matter and where PBH abundances are in consequence likely to be enhanced, as potential  sources of EMRI signals for experiments such as LISA? We briefly discuss such open theoretical and observational  questions below.

\subsection{What is the  abundance of PBHs?}
The formation of a PBH in the early universe is a rare event and knowing the precise formation probability represents, even nowadays,  a challenge. This issue is of fundamental importance as the formation probability is one of the key ingredients to calculate the current PBH abundance and mass function which  not only enter in many observables,  such as the merger rate of BH binaries, but which are also routinely used to express  current constraints from various observations.

To illustrate the difficulty in assessing the precise abundance of PBHs, let us consider the most standard scenario in the literature where the PBHs are formed by the collapse of large inhomogeneities generated during the inflationary stage. The latter delivers a stochastic quantity, the curvature perturbation $\zeta$ on superhorizon scales, whose  properties are known statistically. As we mentioned, PBHs are rare events which require a given observable, e.g. the density contrast $\delta$ or the compaction function $C$,  to be above a typically large  critical value. This implies that the calculation of the PBH abundance requires  going beyond linear perturbation theory and the knowledge of   the tail of the probability where  the latter is typically far  from being a simple Gaussian distribution. Nonlinearities enter into the game even when the initial curvature perturbation is Gaussian (which is typically not) because of the nonlinear relation between the overdensity/compaction function an the curvature perturbation itself. The logic chain is (in radiation)

\begin{equation}
 \textrm{non-Gaussian}\, \zeta(r)\, \rightarrow 
\textrm{non-Gaussian compaction function}\,C(r)=C_1(r)-\frac{3}{8}C_1^2(r),\,\,\,C_1=-\frac{4}{3}r\zeta'(r).
\end{equation}
The presence of nonlinearities at the various stages of the calculation makes  the calculation cumbersome.
One more complication comes from the fact that  the critical threshold $C_c$ to form a PBH depends on the shape of the initial perturbation, specifically on the curvature around the maximum $r_m$ of the compaction function

\begin{equation}
C_c=C_c(q),\,\,\,\,  q=-\frac{r_m^2C''(r_m)}{4C(r_m)},  
\end{equation}
where primes indicate derivatives with respect to the radius and we are assuming spherical symmetry. The formation probability therefore
needs the knowledge of the exact probability $P[C(r_m),C''(r_m)]$.
 Given an initial profile of the curvature perturbation, one neeeds therefore to  compute the corresponding critical profile for the compaction function. The latter determines the position of the maximum $r_m$ in units of the typical momentum scale $k_\star$ of the power spectrum of the curvature perturbation and the correlators derived from the compaction function,  which in turn determine its non-Gaussian statistics. Even though the dependence on the profile may be accounted for integrating over all possible values of the curvature of the compaction function at its maximum in real space or realizing that in some cases the broadest profiles count \cite{Ianniccari:2024bkh},  the position of the maximum $r_m$ is typically calculated only for the average profile of the curvature perturbation and this automatically introduces some intrinsic and unavoidable uncertainties due to the statistical nature of the observables. A possible, albeit incomplete, way to proceed might be to find an observable whose critical value  does not depend on its profile, e.g. the volume average of the compaction function, as suggested in Ref. \cite{Escriva:2019phb} 
 
 \begin{equation}
 \overline C(r_m)=\frac{3}{R_m^3}\int_0^{R_m}{\rm d} x\, x^2\, C(x),\,\,\, R_m=r_m\, e^{\zeta(r_m)},    
 \end{equation}
 but no calculation exists in the literature so far and no understanding of why the volume average of the compaction function has a critical threshold which does not depend on the profile.
Even without   the need of mentioning other possible problems, e.g. dealing with the nonlinearities entering in the radiation transfer function when perturbations re-enter the horizon \cite{DeLuca:2023tun} or the need to deal with the phenomenon of operator mixing \cite{Franciolini:2023wun},  we believe the calculation of the PBH abundance in the standard scenario is still an open issue which deserves further study. The same is true in alternative mechanisms to generate the PBHs, e.g. in  supercooled first-order phase transitions \cite{Flores:2024lng} where the threshold criterion does not suffice.

\subsection{What is the effect of   PBH clustering?} 
Another open issue worth mentioning is the clustering of PBHs. 
It  may significantly affect the merger rates of coalescing binaries and, consequently, the gravitational wave signal measured by the LIGO/Virgo collaboration and by future experiments. 
  Indeed, even if  isolated PBH binaries are not significantly affected by  encounters with a third PBH, PBH binaries residing in PBH clusters might suffer  such interactions,  thus modifying  the expected merger rate.  If  PBHs cluster or not is therefore  an important question to address. 
  A significant PBH clustering might also modify constraints arising from microlensing observations and from the cosmic microwave background. Clustering might also be relevant in determining PBH spins. While the gravitational collapse of a spherical overdensity during radiation domination generates nearly spinless PBHs \cite{DeLuca:2019buf}, they may acquire a large spin in the presence of accretion \cite{DeLuca:2020fpg}. Indeed, if  PBHs cluster they can more easily merge to form binaries at late times, so that the   resulting PBH coming from such  merger  will have  a non-zero value even when no accretion is present.

Even though PBHs in the standard scenario formed  through the collapse of large overdensities  are initially not clustered \cite{Ali_Ha_moud_2018,Desjacques:2018wuu}, PBH clusters form not long after matter-radiation equality \cite{Inman:2019wvr} if $f_{\rm PBH}$ is large. Current N-body simulations are limited to large redshifts $z\sim 10^2$ and a  chief challenge is how to  extrapolate  to lower redshifts relevant for some observational constraints. Analytical   insight is certainly called for  in understanding the complex evolution of the PBH population.

PBHs, even if rare, are potentially important for seeding galaxy and even quasar formation at very high redshift \cite{Carr_2018}. These would be intermediate mass PBHs that enhanced local growth by gravitational instability of cold dark matter. in such a scenario, both massive galaxies and SMBH would form at epochs much earlier than in the standard model of structure formation. Moreover SMBH formation would precede or be contemporaneous with massive galaxy formation, as actually favored by recent JWST observations  \cite{silk2024came}.

\subsection{What fraction of the currently observed GW events can be
ascribed to PBHs?}
Thanks to the many  binary BH events detected so far with many more expected in the next few years, gravitational-wave astronomy is   about to become  a  population study. Currently  hierarchical Bayesian analysis  combining  astrophysical formation models and PBHs may only  constrain the fraction of a putative subpopulation of PBHs in the data (see for instance Ref. \cite{Franciolini:2021tla}). The exciting possibility that PBHs are  contributing to current  observations will   be verified only by further reducing uncertainties in astrophysical and primordial formation models. This is another fundamental open issue.

\subsection{Are PBHs the  Dark Matter?}
The idea that PBH can comprise most of the dark matter is one of the main motivations for studying PBH.  Unfortunately, observational constraints eliminate this   possibility for most of the range of possible  PBH masses, with a notable exception around asteroid mass PBH,  spanning several decades in mass around $\sim 10^{-12}\,  \rm M_\odot$ down to the  limit $\sim 10^{-16}\,  \rm M_\odot$, obtained from the isotropic x-ray and soft gamma ray background observational limits on fluxes   produced by  PBHs currently undergoing Hawking evaporation \cite{Iguaz_2021}.

In the standard formation scenario of PBH,   it is unavoidable that gravitational waves  are generated with a  frequency which  today is in the  mHz range,  exactly  where the LISA mission has maximum sensitivity \cite{Bartolo:2018rku}. The  scenario of PBH as dark matter can be therefore  tested in the future by LISA by measuring the GW two-point correlator. The fact that the asteroid mass range is still unconstrained is due to the fact that the  microlensing  constraints are ineffective   around the value $10^{-11}\, M_\odot$ under which the geometric optics approximation is no longer  valid and the constraints from the presence of neutron stars in globular clusters is based on extreme assumptions about the dark matter density. It is of tantamount importance to come out with possible ideas for constraining or identifying PBH in the asteroid mass range.

One promising approach is that PBH capture leads to neutron star conversion to BH. This could occur in dense star clusters that contain DM, as may be the case for nuclear star clusters, and NS conversion would occur  in the case of capture of such  "endoparasitic" PBHs  for PBH masses larger than $\sim 10^{-11} \, M_\odot$ \cite{Richards_2021}. Such a phenomenon  may produce a deficit of pulsars in our GC as possibly observed \cite{Dexter_2014}. Another interesting avenue is to observe the  number of  massive main-sequence stars in ultra-faint dwarfs which should be suppressed if the dark matter is  made of asteroid-mass PBHs \cite{Esser:2023yut}, making measured the mass distribution of stars  depleted in the high-mass range. 

Furthermore, near-extremal PBHs provide an intriguing means of making evaporating PBHs stable on cosmological scales. Formation is simplest for low mass five-dimensional PBHs that initially act like four-dimensional PBHs,  Hawking radiating down to the radius of the extra dimension where their effective temperature is effectively zero, to attain a stable mass \cite{Pacheco_2023}. These are  generated in higher dimensional scenarios and Hawking radiation is generically found to be slowed down \cite{2022PhRvD.106h6001A}. Other scenarios for producing  near-extremal PBH include formation of  maximally rotating or charged PBH at very early epochs  \cite{deFreitasPacheco:2020wdg}, as well as via the quantum gravity phenomenon of so-called   memory burden suppression \cite{Alexandre:2024nuo}. Detectability is feasible  down to  scales as small as a few Planck masses  for charged PBH relics via terrestrial detectors \cite{Lehmann_2019} or equally  for high energy particle emission from occasional binary merger events  \cite{Bai_2020}.

\section{The PBH Roadmap}
Future experiments abound, ranging from approved projects to imminently enter the detailed design phase, to provide  significantly enhanced constraints on PBH scenarios, including  terrestrial laser interferometers (Einstein Telescope, Cosmic Explorer)\cite{domènech2024probing}, and  space laser interferometers (LISA, DECIGO \cite{Kawamura:2020pcg}, TianQin and Taiji \cite{Gong_2021}), to atomic interferometers on the ground with demonstrators under  construction  and  extensions planned  to km scales (AION MAGIS, MAGIA, MIGA, ELGA, ZAIGA),  as well as in space (AEDGE) \cite{ellis2023probing},  and designs for lunar interferometer/seismometer arrays (LGWA, GLOC)\cite{van_Heijningen_2023}. 

The science goals focus on bridging frequency gaps in the LIGO/VIRGO/KAGRA frequency range that has so far led to hundreds of detections of BH mergers in the $(10-100) M_\odot$ range by going both to higher and lower frequencies. The new science that can be explored ranges from the masses and equation of state of neutron stars as well as multimessenger signals from core collapse supernovae to kpc distances  (kHz to Hz range) to SMBH and IMBH detections and BH inspiralling (Hz to mHz,  and eventually  $\mu$Hz). One would ultimately approach the  pulsar timing array sensitivities (NANOGrav, SKA) at nanohertz frequencies to enable exploration of first order very early universe phase transitions,  and presence of loop decays in a network of  cosmic strings \cite{ellis2023source}.

We next address  various ``smoking-gun" signatures that may identify  PBHs, notably including  high redshift mergers and  sub-solar mass PBHs.

\subsection{High redshift mergers}
The PBH model predicts a binary merger rate density which grows monotonically with redshift~\cite{Ali-Haimoud:2017rtz,Raidal:2018bbj,DeLuca:2020qqa}. Focusing those binaries generated at early epochs, one has a well-defined time evolution with redshift
\begin{equation}
    R^{\rm EB} (z) \propto\left(\frac{t(z)}{t (z=0)}\right)^{-34/37},
\end{equation} 
extending up to redshifts $z\gtsim{\cal O}(10^3)$.  
Since   mergers from BHs of astrophysical origin switch off at  redshifts $z\gtsim (20-30)$~\cite{Kinugawa:2014zha,Kinugawa:2015nla,Hartwig:2016nde,Belczynski:2016ieo,Inayoshi:2017mrs,Liu:2020lmi,Liu:2020ufc,Kinugawa:2020ego,Tanikawa:2020cca,Singh:2021zah}, detecting merger redshifts at $z\gtsim30$ will  represent smoking guns for primordial binaries~\cite{Koushiappas:2017kqm,DeLuca:2021wjr,Ng:2021sqn}. Of course, large measurement uncertainties  on the inferred luminosity distance due to the low SNR at 3G detectors \cite{Ng:2021sqn,Franciolini:2021xbq,Martinelli:2022elq,Ng:2022vbz} may bring relatively large errors, but recent studies show that identifying  merger rate evolution at redshift larger than ${\cal O}(10)$ will 
 allow to constrain PBH populations up to abundances as low as $f_{\rm PBH} \approx 10^{-5}$ \cite{Ng:2022agi} in the solar mass range, even accounting for the contamination of Pop III binaries, see Fig. 1.
 \begin{figure}[hbt]
\centering
  \includegraphics[width=10cm]{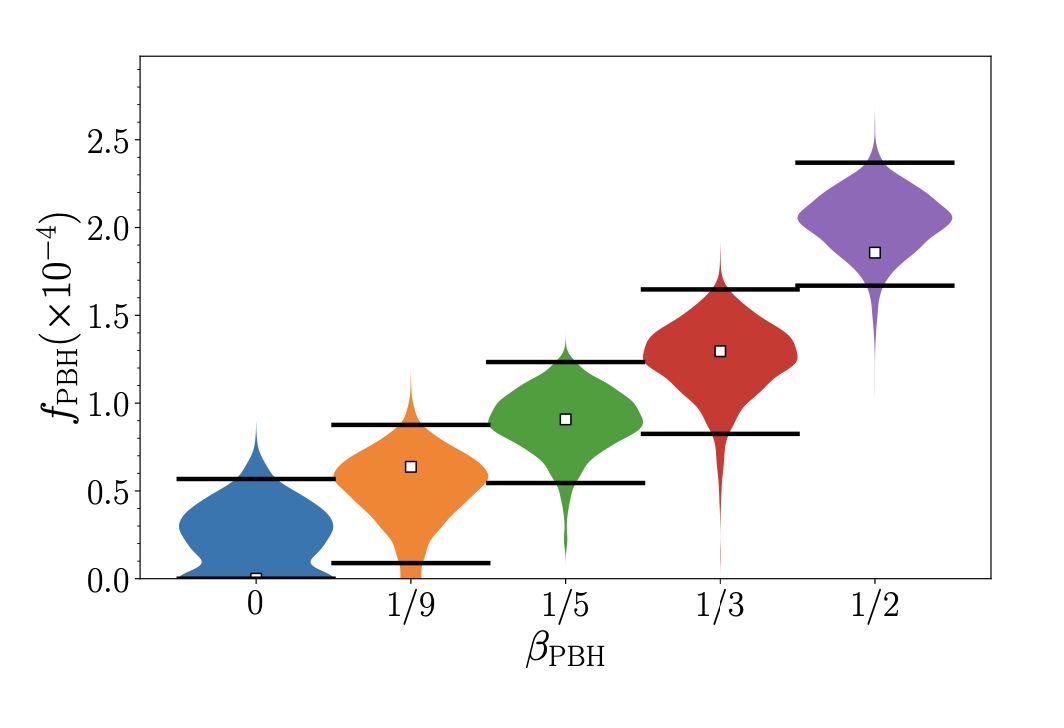}
  \caption{Posteriors of $f_{\rm PBH}$ given a lognormal PBH mass function centered at  $M_c=30\, M_\odot$ and variance equal to 0.3 
  at $\beta_{\rm PBH} = 0, 1/9, 1/5, 1/3$, and $1/2$, being $\beta_{\rm PBH}$ the fraction of the PBH merger rate to the total merger rate.  The solid black lines indicate the 95\% CIs. From Ref. \cite{Ng:2022agi}.}
\end{figure}

\subsection{Sub-solar PBHs}
The detection of  sub-solar BHs 
in a compact binary merger is regarded as one of the smoking gun signatures of  PBHs. Indeed, BHs of astrophysical origin are not 
 expected. The sub-solar mass range may be however  populated by 
other compact objects like neutron stars, white dwarfs, or exotic compact objects~\cite{Cardoso:2019rvt} (e.g. boson stars~\cite{Guo:2019sns}).

GW signatures of any  population of PBH below the astrophysical production provide a powerful statistical constraint on any solar or subsolar component, as evidenced by the sharp decline observed  in 03 events. There could of course be rare events generated by solar or subsolar PBHs that might contribute to DM at the ${\cal O}(10\%)$ level.
One key ingredient is therefore  to distinguish PBHs from other compact objects, e.g.  through the measurements of the  tidal disruption and tidal deformability $\Lambda$. A recent analysis has shown that 
for subsolar neutron-star binaries, the O4 and O5 projected sensitivities would allow measuring the effect of tidal disruption on the waveform in a large portion of the parameter space, also constraining the tidal deformability at ${\cal O}(10\%)$ level, thus excluding a primordial origin of the binary.
Viceversa, for subsolar PBH binaries, model-agnostic upper bounds on the tidal deformability can rule out neutron stars or more exotic competitors.
Assuming events similar to the sub-threshold candidate SSM200308 reported in LVK O3b data are PBH binaries, O4 projected sensitivity would allow ruling out the presence of neutron-star tidal effects at $\approx 3 \sigma$ C.L., thus strengthening the PBH hypothesis. Future experiments would lead to even stronger ($>5\sigma$) conclusions on potential discoveries of this kind, see Fig. 2. 

 PBHs of lunar mass, a regime where hints of  unaccounted-for microlensing events have been reported by the OGLE survey of the galactic bulge  \cite{Niikura_2019}, could be sufficiently numerous that nearby binaries might be detectable with future generation detectors \cite{Inoue_2003}.
\begin{figure}[hbt]
\centering
  \includegraphics[width=10cm]{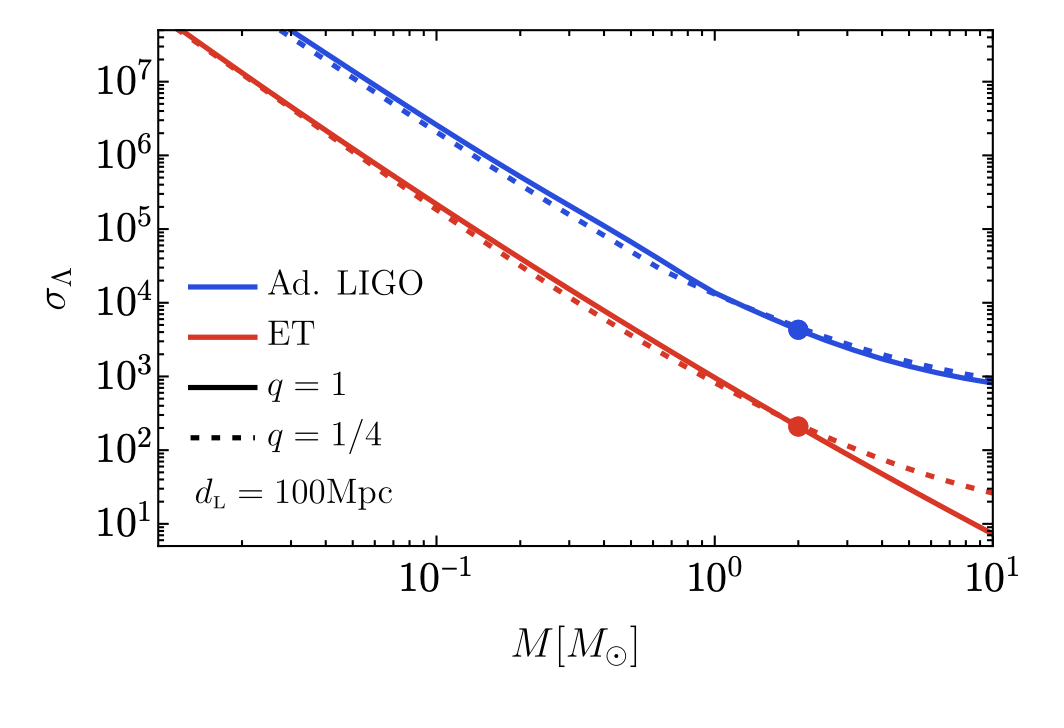}
  \caption{Precision of the deformability parameter $\Lambda$ at both Ad. LIGO and ET. The binary is assumed to have spinless components and negligible eccentricity and deformability, as predicted by the PBH scenario. The solid (dashed) line indicates the result for mass ratios 1 $(1/4)$. From Ref. \cite{Franciolini:2021xbq}.}
\end{figure}
\subsection{Plugging the pair instability gap  with PBH?}
An equally powerful mass challenge for  PBH arises from the pair instability gap. This is generated by  the catastrophic core collapse of massive stars that results from  pair-instability SNe,  and predicts a deficit of astrophysical  BH between  $\sim 50\, M_\odot$ and $\sim 120\, M_\odot.$ The full LIGO-Virgo O3 data set reveals a definitive decline in merger rates above $\sim 40\, M_\odot$ \cite{theligoscientificcollaboration2022population} but with 
several candidates filling this gap at high credibility \cite{wadekar2023new}.

Could PBH populate  the pair instability  gap? A smoking gun signature would be a feature in the BH mass spectrum, most likely indicative of a new formation mechanism. 

\subsection{PBH eccentricity}
Even though PBH binaries are  formed with large eccentricity at high redshift, they have   enough time to circularize before the GW signal can enter the observation band of current and future detectors. 
This implies that an observation of  a non-zero eccentricity $e$ would rule out the interpretation as a primordial binary formed in the early Universe, while it may still be compatible with a PBH binary formed in the late-time universe \cite{Cholis:2016kqi,Wang:2021qsu}. 

\subsection{PBH spin}
The  PBHs form in the early universe through the standard scenario of the collapse of large overdensities inherit  a  characteristic mass-spin correlations induced by accretion effects \cite{DeLuca:2020fpg}.
Using this criterion for determining the possible primordial nature of individual GW events would require reducing uncertainties on the accretion model and studied at the  population level \cite{Franciolini:2022iaa}.

Fig. 3 presents a flowchart from Ref. \cite{Franciolini:2021xbq} one might follow to  assess the primordial nature of a binary merger based on measurements of the redshift $z$, eccentricity $e$, tidal deformability $\Lambda$, component masses $m$, and dimensionless spins $\chi$.

\begin{figure}[hbt]
\centering
  \includegraphics[width=13cm]{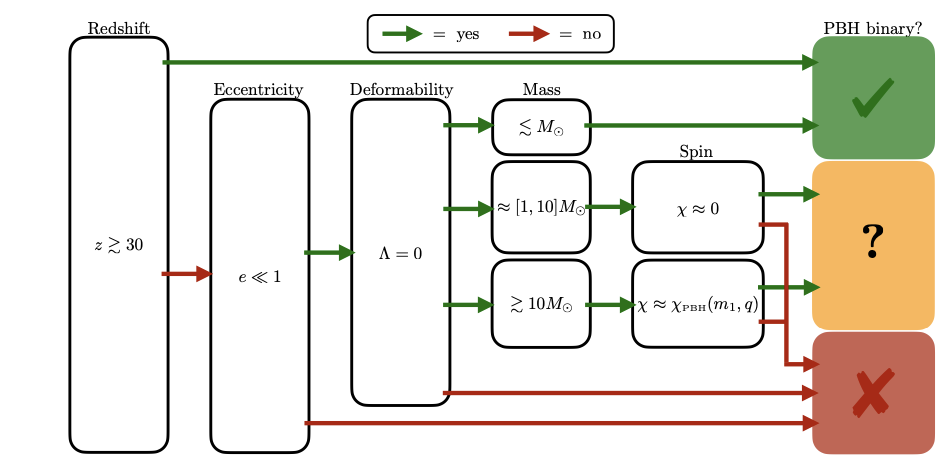}
  \caption{Flowchart from Ref. \cite{Franciolini:2021xbq} to  assess the primordial nature of a binary merger based on measurements of the redshift $z$, eccentricity $e$, tidal deformability $\Lambda$, component masses $m$, and dimensionless spins $\chi$. Each arrow indicates if the condition in the box is met (green) or violated (red).}
\end{figure}
\noindent



\subsection{Future gamma-ray telescopes}
PBHs have a range of potentially interesting direct gamma ray signals associated with Hawking evaporation.  These include  the 511 keV gamma-ray line, produced by electron-positron pair-annihilation, where positrons originate from black hole evaporation. The INTEGRAL detection of the Large Magellanic Cloud provides one of the strongest bounds attainable with present observations, and should be greatly improved upon by  future MeV gamma-ray telescopes such as GECCO \cite{Korwar_2023}, as well as AMEGO \cite{fleischhack2021amegox} and  ASTROGRAM \cite{Mallamaci:20198W}, among others \cite{Bird_2023}.

Exploding PBHs, the late-time limit of Hawking evaporation for small PBHs, provide intriguing candidates for future very high energy gamma ray telescopes that may address transient phenomena, that pertain especially to the new frontier of PBHs of ultralralow masses. 

The Hawking temperature characterizes the effective temperature as inversely proportional to PBH mass and is constrained as discussed previously by soft gamma ray constraints at the $\sim 10^{17}\rm gm$ PBH mass  scale, incorporating appropriate spectral corrections due to nonthermal aspects of the complex emission processes \cite{Auffinger_2023}.  Once the  allowable PBH mass range is extended to much lower masses due to slowing or suppression of Hawking radiation, as proposed in recent discussions of near-extremal and higher dimensional PBHs, intriguing prospects arise for high energy gamma ray experiments 
\cite{deFreitasPacheco:2020wdg, 2022PhRvD.106h6001A, 1940PCPS...36..325H, Alexandre:2024nuo}. As previously noted, new observational terrain opens up  
for exploring near-extremal PBH  detection both directly via terrestrial experiments \cite{Lehmann_2019} and indirectly via  binary merger events \cite{Bai_2020},
where constraints at much higher gamma ray or even neutrino energies could become important.  

Such gamma ray   experiments include HAWC (100 GeV -50 TeV) and CTA (20 GeV-300 TeV) \cite{boluna2024searching} at TeV scales, and at PeV-scale photon energies LHAASO \cite{LHAASO:2021gok} and SWGO \cite{2022icrc.confE..23H}. Such searches will be closely coordinated with multimessenger projects spanning next generation underground gravitational wave telescopes  (EINSTEIN (ET), COSMIC EXPLORER  (CE)) and FRB observatories 
(CHORD) \cite{https://doi.org/10.5281/zenodo.3765414}.

\section{Conclusions}
Support for the existence of PBHs  hypothesis comes from a range of arguments spanning dark matter physics, inflationary models, astrophysics, gravitational microlensing   and quantum cosmology. None are yet conclusive but the cumulative support for the existence of PBHs seems reasonably compelling.  More to the point however is that there is a rich astrophysical agenda  of open questions that contribute to a future road map for a plethora of research projects.

Two of the most fundamental PBH probes that distinguish them from astrophysical counterparts are their redshift and mass distributions.This combination provides an armoury of  smoking guns that will inevitably enrich the next generation of gravitational wave and gravitational microlensing experiments. Confirmation will require improved statistics as single  events may not be conclusive. it is only too easy to come  up with rare events that  can masquerade as PBH candidates, for example by appealing to higher order generations of merging events or to non-Gaussianity. We have laid out a possible future road map for definitively assessing whether PBHs can present a viable candidate for dark matter, or possibly  contribute to other astrophysical or even quantum cosmology anomalies.

\bibliographystyle{unsrt}
\bibliography{main}

\end{document}